\documentclass[final,1p,times]{elsarticle}
\usepackage{graphicx}
\usepackage{graphics}
\usepackage{amssymb}

\begin{document}

\begin{frontmatter}

\title{A discerning gravitational property for gravitational equation in higher dimensions}

\author[nkd1,nkd2]{Naresh Dadhich}
\ead{nkd@iucaa.ernet.in}



\address[nkd1]{Centre for Theoretical Physics, Jamia Millia Islamia,
New Delhi 110025, India}

\address[nkd2]{Inter-University Centre for
Astronomy \& Astrophysics, Post Bag 4, Pune 411 007, India}


\begin{abstract}
It is well-known that Einstein gravity is kinematic (no non-trivial vacuum solution;i.e. Riemann vanishes whenever Ricci does so) in $3$ dimension because Riemann
 is entirely given in terms of Ricci. Could this property be universalized for all odd dimensions in a generalized theory? The answer is yes, and this property uniquely singles out pure Lovelock (it has only one $N$th order term in action) gravity for which $N$th order Lovelock Riemann tensor is indeed  given in terms of corresponding Ricci for all odd $d=2N+1$ dimensions. This feature of gravity is realized only in higher dimensions and it uniquely picks out pure Lovelock gravity from all other generalizations of Einstein gravity. It serves as a good discerning and guiding criterion for gravitational equation in higher dimensions.                              
\end{abstract}

\end{frontmatter}


\medskip

\section{Introduction}

Absence of all forces is characterized by maximally symmetric spacetime of constant (homogeneous) curvature, and Einstein gravity (GR) naturally arises when spacetime turns inhomogeneous \cite{measure-grav}. It is the Riemann curvature that should determine dynamics of the force responsible for its inhomogeneity.    The Riemann curvature satisfies the Bianchi differential identity which is purely a differential geometric property and its trace yields a second rank symmetric tensor with vanishing divergence, the Einstein tensor, giving the second order differential operator for the equation of motion. This is how we are uniquely led to the Einstein gravitational equation on identifying the cause of inhomogeneity as matter-energy, a universal physical property for all that physically exist, distribution \cite{measure-grav}. Thus gravitational dynamics is entirely determined by sometime curvature and it resides in it. \\

It is very illuminating that without asking for an equation for gravity, GR simply follows from the geometric properties of Riemann curvature. Similarly, does geometry also determine spacetime dimension? The second order differential operator in the equation is given by the Einstein tensor which is non-trivial only in dimensions $d>2$. Next the equation should admit non-trivial vacuum solution for free propagation which requires $d>3$. In $d=3$, Riemann curvature is entirely given in terms of Ricci tensor;i.e. it vanishes whenever Ricci vanishes. That is how gravity is kinematic in $d=3$ as vacuum is flat and absence of non-trivial vacuum solution signifies absence of free degrees of freedom for propagation of the field. This is how we come to the usual four dimensional spacetime that admits non-trivial vacuum solution. However the Einstein equation would be valid in all higher dimensions as well. \\

We could however ask the question, could the kinematic property of GR for odd $d=3$ dimension be universalized to all odd dimensions? Naturally this would require a generalization of GR because GR could be kinematic only in three dimension and none else. However this gravitational property may serve as a good guiding principle for a gravitational equation in higher dimensions. That is what we shall probe and establish that it uniquely singles out pure Lovelock gravity for which action consists of only one $N$th order term of the Lovelock polynomial Lagrangian. Lovelock is the most natural generalization of GR because it is the only one that remarkably retains the second order character of the equation despite the action being a homogeneous polynomial in Riemann. This is quintessentially a higher dimensional generalization of GR. \\

In this essay we shall proceed as follows: First, we shall establish that gravity is indeed kinematic \cite{dgj, xd} in all odd $d=2N+1$ dimensions in pure Lovelock gravity relative to properly defined Lovelock analogues of Riemann and Ricci tensors \cite{bianchi, kastor, xd}. Since this generalization is effective only in higher dimensions, it is therefore pertinent to ask what is it that demands higher dimension(s), and why is it small so that it is not accessible to present day observations? That is what we shall probe next by appealing to a general principle and some gravitational properties. Next we would argue that pure Lovelock is thus proper gravitational equation \cite{eqn} in higher dimensions and it is the only one that obeys kinematicity of gravity in all odd $d=2N+1$ dimensions. It is valid only for two, odd $d=2N+1$ and even $d=2N+2$, dimensions. It includes GR in the linear order $N=1$. We would wind up the discourse with a discussion. \\

\section{Kinematic property}

Dadhich \cite{bianchi} defined an appropriate Lovelock analogue of Riemann, which was a homogeneous polynomial in Riemann, with the property that trace of its Bianchi derivative vanished. That gave rise to a corresponding analogue of Einstein tensor which was the same as the one obtained by varying the corresponding Lovelock action. Using this Lovelock Riemann generalization it was first shown that static vacuum spacetime was kinematic in all odd $d=2N+1$ dimensions;i.e. vanishing of Ricci implied vanishing of Riemann \cite{dgj}. Right on the heels of this discovery came yet another parallel definition of Lovelock Riemann tensor by Kastor \cite{kastor} which involved a $(2N,2N)$-rank tensor, completely antisymmetric both in its upper and lower indices. Though the two Lovelock Riemann analogues are not completely equivalent yet interestingly they both yield the same Einstein tensor, and hence the same equation of motion. The difference between them came to the fore recently while studying vacuum solutions for the Kasner metric \cite{dmx} where kinematic property held good for the Kastor Lovelock Riemann but not for the Dadhich one. Interestingly for static spacetime, the difference between the two vanishes and that is why it was not noticed earlier. \\

In view of its general validity, we shall employ the Kastor Lovelock Riemann tensor for establishing the kinematic property in all odd $d=2N+1$ dimensions. This ($2N,2N$)-rank tensor is defined as follows \cite{kastor}:

\begin{equation}
\mathbb{R}^{b_1 b_2 \cdots b_{2N}}_{a_1 a_2 \cdots a_{2N}}= R^{[b_1 b_2}_{\quad \quad [a_1 a_2}\cdots R^{b_{2N-1} b_{2N}]}_{\qquad \qquad a_{2N-1} a_{2N}]}~.
\label{kastensor}
\end{equation}
It is product of $N$ Riemann tensors and completely antisymmetric in both its upper and lower indices. With all indices lowered, it is also symmetric under exchange of both groups of indices, $a_i\leftrightarrow b_i$.

The Lovelock Lagrangian is written as
\begin{equation}
\left.\right.\mathcal{L}= \frac{2^N}{(2N)!(d-2N)!}\epsilon_{a_1 a_2\cdots a_d}\,\mathbb{R}^{a_1a_2\cdots a_{2N}}\wedge e^{a_{2N+1}}\wedge \cdots \wedge e^{a_d} ~,\\
\end{equation}

giving rise to the corresponding Einstein tensor

\begin{equation}
\left.\right.\mathcal{E}_{\ c}^{b}= \frac{2^N}{(2N)!(d-2N-1)!}\,\epsilon_{a_1 a_2\cdots a_{d-1} c}\,\mathbb{R}^{a_1a_2\cdots a_{2N}}\wedge e^{a_{2N+1}}\wedge \cdots \wedge e^{a_{d-1}}\wedge e^b ~.
\label{einstein}
\end{equation}

It is purely an algebraic property that for $d=2N+1$, the above defined $4N$th Lovelock Riemann tensor could be entirely written in terms of its contraction Ricci, and thereby Einstein tensor, and it is in fact written as follows \cite{xd},
\begin{equation}
\mathbb{R}^{b_1\cdots b_{2N}}_{a_1\cdots a_{2N}}=\frac{1}{(2N)!}\epsilon^{b_1\cdots b_{2N+1}}\,\epsilon_{a_1\cdots a_{2N+1}}\mathcal{E}^{a_{2N+1}}_{\qquad b_{2N+1}}~.
\label{main}
\end{equation}

This clearly establishes the kinematic property that Lovelock Riemann vanishes in all odd $d=2N+1$ dimensions whenever the corresponding Einstein (Ricci) tensor vanishes. It may however be mentioned that though vacuum spacetime in odd dimensions would be Lovelock Riemann flat but it would not in general be Riemann flat \cite{dgj}. Another way of characterizing kinematic property is that the corresponding Weyl curvature vanishes in all odd $d=2N+1$ dimensions. \\

\section{Why (small) higher dimensions?}

It is symmetries of field theory for a consistent theory of fundamental particles and their interactions that lead naturally to higher dimensions, and this paradigm is popularly known as string theory. It is all driven by field theoretic considerations without any direct reference to gravity. Instead I would here like to concern myself only to gravitation to ask the question, are there any gravitational features that have so far remained unaddressed and whether their inclusion asks for higher dimensions? \\

One such possible feature could be probing of gravity in high energy regime \cite{measure-grav, higher-d}. For addressing high energy effects of any theory, we generally include higher powers of the basic field entity which in the present case is Riemann tensor. In Einstein gravity, Riemann occurs linearly in action, for high energy considerations we should therefore include higher powers of Riemann in action. However at the same time if we demand that the basic character of the equation should not change;i.e. it should continue to remain second order differential equation which is also required for warding off undesirable features like occurrence of ghosts. This uniquely singles out the Lovelock action which alone has this remarkable property that the equation continues to remain second order despite action being homogeneous polynomial in Riemann tensor. But higher order Riemann terms in Lovelock action make non-zero contribution to the equation only in dimension $>4$. That is why higher dimensions are required to probe high energy effects of gravity \cite{higher-d}. \\

Second we appeal to the general principle that total charge of a classical field must be zero when it is summed over all charges in the universe.
It is obviously true for electric field because charge is created from a neutral entity like an atom by kicking out one polarity particle out and what remains behind has equal and opposite charge. When all charges are summed over, it will add to zero. This should also happen for gravity. The charge for gravity is matter-energy distribution which has only one polarity, always positive by convention. How could this be balanced as there exists no matter-energy of opposite polarity, but balance it must to obey the general principle of total charge being zero. The only way out is that gravitational field that it produces must itself have charge of opposite polarity;i.e. gravitational field energy must be negative. This is why gravity can only be attractive and it is dictated by this general principle. Though negative gravitational charge is non-localizable as it is spread over whole of space, if we integrate it over entire space it would completely  balance positive charge of matter distribution. This is exactly what has been rigorously shown in the famous ADM paper \cite{adm}. \\

Now consider a $3$-ball of some finite radius around a mass point. Since total charge on the ball is non-zero, because negative charge in the field lying outside the ball has been cut off, field must therefore propagate of the ball in higher dimension. However as it propagates, its past lightcone would include the region, and hence negative charge of the field, lying outside. Thus propagation in extra dimension would not be free but instead be with diminishing field strength (or equivalently it could be viewed as massive propagation). Thus gravity propagates in higher dimension but not deep enough \cite{higher-d}. This is precisely what the Randall-Sundrum braneworld gravity \cite{rand-sund} envisages where the usual massless free propagation remains confined to the brane while off the brane propagation is massive. The braneworld model is string theory inspired while ours is a purely classical consideration based on a very general principle.  \\

These two are purely classical arguments that appeal to general principles and considerations for a classical field. They clearly point to the fact that gravity could not remain confined entirely to a particular dimension. This is because gravitational dynamics resides in spacetime curvature and hence it cannot be constrained by any external prescription. Gravity is thus entirely self driven. This is the critical property that distinguishes gravity from all other forces. This is why only gravity propagates in higher dimensions while all other matter fields are believed to remain confined to the usual four dimensions \cite{why-fourd}. Existence or realization of spacetime dimension is probed by observing propagation of a field in that. Since only gravity can propagate in higher dimension but not deep enough, hence higher dimension(s) cannot be large. The application of the general principle of total charge being zero not only asks for higher dimension, but also prescribes it to be small. This is a remarkable conclusion following from a very general principle. \\

\section{Gravitational equation in higher dimensions}

Since gravity cannot remain entirely confined to a given dimension, the consideration of higher dimensions becomes pertinent, and then arises the question what should be the equation of motion in higher dimensions? Note that we are here not seeking an effective equation that takes into account some semi-classical corrections instead we are asking for a classical equation in higher dimensions. For that, the first and foremost requirement is that it should be of second order which uniquely picks out the Lovelock polynomial action in which each term comes with dimensionful coupling constant. Also note that Lovelock Lagrangian is the most general invariant that can be constructed from Riemann tensor giving the second order equation of motion. \\

On the other hand one can carry on with the Einstein equation itself which is valid in all dimensions $d\geq3$. This would however not be the most general equation in dimensions $>4$ while the Lovelock polynomial action would give the most general equation for all $d\geq2N+1$,  and it includes GR for $N=1$. The problem with the Lovelock equation is that it has a dimensionful coupling for each $N$, and there is no way to determine more than one coupling by measuring the strength of the field, which is only one. Thus there is arbitrariness in fixing couplings. \\

There is one way out that by invoking some property of gravity if we can justify that Lovelock polynomial should involve only one $N$th order term. That would then be what we have called pure Lovelock. And that property is indeed universalization of kinematic property;i.e. gravity be kinematic (non-existence of non-trivial vacuum spacetime) in all odd $d=2N+1$ dimensions. Thus kinematic property uniquely picks out pure Lovelock gravity. This equation would be  valid only for two (odd and even, $d=2N+1, 2N+2$) dimensions because else kinematic property will be violated. \\

Note that pure Lovelock equation \cite{eqn} has several interesting and desirable features. For instance even though the equation is completely free of the Einstein term yet static vacuum solution with $\Lambda$ asymptotically goes over to Einstein-dS solution in the given dimension \cite{dpp-sol}. This is quite remarkable that pure Lovelock solution includes Einstein gravity asymptotically even though the equation is  completely free of it. Similarly, bound orbits around static black hole exist in pure Lovelock gravity in all even $d=2N+2$ dimensions in contrast for Einstein gravity they do  only in $4$ dimension and none else \cite{dgj-bound}. Also thermodynamical parameters, temperature and entropy bear universal relation with horizon radius in all odd and even $d=2N+1, 2N+2$ dimensions, in particular entropy always goes as square of horizon radius in all even dimensions \cite{dpp-thermo}. \\

Thus pure Lovelock equation has all the features what one would have asked for a gravitational equation. The newly recognised kinematic property is clearly its distinguishing feature. It is thus the right gravitational equation \cite{eqn} in higher dimensions $d=2N+1, 2N+2$. That is for each $N$, the equation is only for the corresponding two odd and even dimensions, for instance for N=1, Einstein equation is only for $d=3, 4$, for $N=2$, pure GB equation only for $d=5, 6$, and so on. The Einstein equation is therefore good only for three and four dimensions, and in higher dimensions we should go over to next order of $N$. Thus pure Lovelock gravity is a new paradigm for higher dimensions, and it is the kinematic property that has played the key discerning role. \\

\section{Discussion}

By appealing to universalization of kinematic property (that there exists no non-trivial vacuum solution in odd dimensions) for all odd dimensions, we have arrived at the unique gravitational equation which is pure Lovelock involving only one $N$th order term. Thus kinematic property plays the key role as discerning criterion as well as a guiding principle for gravitational dynamics in higher dimensions. We have just universalized the already existing property in Einstein gravity to get to the proper equation in higher dimensions. A good and enlightening generalization of a theory always stems from extending some key property beyond the normal premise of the theory, and then the existing theory gets automatically included in the new theory. Pure Lovelock gravity,  which uniquely incorporates kinematic property for all odd $d=2N+1$ dimensions, includes Einstein gravity for $N=1$. \\

Again by appealing to another general principle of total charge being zero for a classical field, we have argued that gravity cannot remain confined to a given dimension instead it propagates off into higher dimension(s) but not deep enough. It is remarkable that this very general consideration has not only asked for higher dimension(s) but also prescribed that it has to be small. \\

The principal aim of the essay is to demonstrate the key discerning role kinematic property plays in picking up the right equation for gravity in higher dimensions. Having done that let's ask what more does it entail? \\

There is famous BTZ black hole solution \cite{btz} in $3$ dimension which is a $\Lambda$-vacuum solution. Note that it is the presence of $\Lambda$ that makes spacetime non-flat.  For Einstein gravity, it can therefore occur only in $3$ dimension and none else. Since pure Lovelock gravity is kinematic in all odd dimensions, hence analogues of BTZ black hole would exist in all odd $d=2N+1$ dimensions \cite{dgj}. \\

All this is very fine, however the key question remains, how does the higher dimensional equation impact on $4$-spacetime we live in? The braneworld model \cite{rand-sund} that envisages propagation of gravity in higher dimension but not deep enough predicts $1/r^3$ correction to the Newtonian potential on the brane corresponding to an  AdS bulk. The situation should be similar in what we are proposing except perhaps we would rather employ pure Gauss-Bonnet equation in the bulk rather than Einstein. This may not however be much relevant so far as AdS (which is a solution of pure GB equation as  well) bulk is concerned. The situation would be different if we consider a pure GB-BTZ black
hole so that Weyl curvature in $5$ dimensional bulk is non-zero which will project down on the brane as tracefree black radiation in the equation. Then we would have black hole on the brane given by Reissner-Nordstrom metric obtained by Dadhich et al \cite{dmpr} where $Q^2$ is not the Maxwell charge but it is Weyl charge appearing in the metric as $-Q^2/r^2$. It is though very insightful, however there exists no complete solution of the bulk-brane system. \\

It is indeed very remarkable and insightful that universalization of certain gravitational property uniquely picks out an equation in higher dimensions. In the same vein, let's further ask, is there any other similar instance of insightful deduction? The one thing that comes to mind is that how should vacuum energy gravitate \cite{measure-grav, cern}? It was argued that vacuum energy was on the same footing as gravitational field energy. Both are created by matter and hence have no independent existence of their own, and therefore they should not gravitate through a stress tensor in the equation notwithstanding whether a stress tensor could be written or not. Clearly we write no stress tensor for gravitational field energy on the right, and in fact it gravitates in a much  subtler manner. It gravitates by enlarging the spacetime framework, by curving $3$-space \cite{ein-new}. That is why Newton's inverse square law remains intact in GR. Something similar should happen for vacuum energy. \\

It is therefore a matter of principle that vacuum energy should not gravitate through a stress tensor but instead by enlargement of framework. This we would not know unless we have quantum theory of gravity. May what that be, $\Lambda$ becomes free of the Planck length and hence it could, as a true constant of spacetime structure \cite{measure-grav},    have any value as being determined by acceleration of the Universe \cite{perl}. Thus it gets liberated and has nothing to do with vacuum energy \cite{measure-grav, cern}. At a concept level this is a very important realization.  \\

The discovery of GR was solely driven by principle and concept, and hence in its  centennial year the present exercise is a fitting tribute to that spirit of doing science and to its great creator. \\

\textbf{Acknowledgement:} The author wishes to thank the Albert Einstein Institute, Golm-Potsdam for a summer visit where the manuscript was completed.

\bibliographystyle{elsarticle-num}

\end{document}